\documentstyle[preprint,prd,aps,amsfonts,epsf]{revtex} 

\begin{document}
\draft
\preprint{\begin{minipage}[t]{3in} 
                \flushright
                FERMILAB-PUB-98/189-E      
        \\      CDF/PUB/BOTTOM/CDFR/4574   
        \\      PRL Version 1.0            
        \\      \today 
        \\  $\;$
                \end{minipage} }

\tightenlines
\title{ 
\boldmath
Measurement of the $CP$-Violation Parameter $\sin(2\beta)$
in $B^0_d/\overline{B}{^0_d} \rightarrow J/\psi K^0_S$ Decays 
}

\def\r#1{\ignorespaces $^{#1}$}
\font\eightit=cmti8

\author{ 
\hfilneg
\begin{sloppypar}
\noindent
F.~Abe,\r {17} H.~Akimoto,\r {39}
A.~Akopian,\r {31} M.~G.~Albrow,\r 7 A.~Amadon,\r 5 S.~R.~Amendolia,\r {27} 
D.~Amidei,\r {20} J.~Antos,\r {33} S.~Aota,\r {37}
G.~Apollinari,\r {31}
% ...
\end{sloppypar}
\vskip .026in
\begin{center}
(CDF Collaboration)
\end{center}
\vskip .026in
\begin{center}
\r 1  {\eightit Argonne National Laboratory, Argonne, Illinois 60439} \\
\r 2  {\eightit Istituto Nazionale di Fisica Nucleare, University of Bologna,
I-40127 Bologna, Italy} \\
\r 3  {\eightit Brandeis University, Waltham, Massachusetts 02254} \\
\r 4  {\eightit University of California at Los Angeles, Los 
Angeles, California  90024} \\  
\r 5  {\eightit University of Chicago, Chicago, Illinois 60637} \\
\r 6  {\eightit Duke University, Durham, North Carolina  27708} \\
\r 7  {\eightit Fermi National Accelerator Laboratory, Batavia, Illinois 
60510} \\
\r 8  {\eightit University of Florida, Gainesville, Florida  32611} \\
\r 9  {\eightit Laboratori Nazionali di Frascati, Istituto Nazionale di Fisica
               Nucleare, I-00044 Frascati, Italy} \\
\r {10} {\eightit University of Geneva, CH-1211 Geneva 4, Switzerland} \\
\r {11} {\eightit Harvard University, Cambridge, Massachusetts 02138} \\
\r {12} {\eightit Hiroshima University, Higashi-Hiroshima 724, Japan} \\
\r {13} {\eightit University of Illinois, Urbana, Illinois 61801} \\
\r {14} {\eightit Institute of Particle Physics, McGill University, Montreal 
H3A 2T8, and University of Toronto,\\ Toronto M5S 1A7, Canada} \\
\r {15} {\eightit The Johns Hopkins University, Baltimore, Maryland 21218} \\
\r {16} {\eightit Institut f\"{u}r Experimentelle Kernphysik, 
Universit\"{a}t Karlsruhe, 76128 Karlsruhe, Germany} \\
\r {17} {\eightit National Laboratory for High Energy Physics (KEK), Tsukuba, 
Ibaraki 305, Japan} \\
\r {18} {\eightit Ernest Orlando Lawrence Berkeley National Laboratory, 
Berkeley, California 94720} \\
\r {19} {\eightit Massachusetts Institute of Technology, Cambridge,
Massachusetts  02139} \\   
\r {20} {\eightit University of Michigan, Ann Arbor, Michigan 48109} \\
\r {21} {\eightit Michigan State University, East Lansing, Michigan  48824} \\
\r {22} {\eightit University of New Mexico, Albuquerque, New Mexico 87131} \\
\r {23} {\eightit The Ohio State University, Columbus, Ohio  43210} \\
\r {24} {\eightit Osaka City University, Osaka 588, Japan} \\
\r {25} {\eightit Universita di Padova, Istituto Nazionale di Fisica 
          Nucleare, Sezione di Padova, I-35131 Padova, Italy} \\
\r {26} {\eightit University of Pennsylvania, Philadelphia, 
        Pennsylvania 19104} \\   
\r {27} {\eightit Istituto Nazionale di Fisica Nucleare, University and Scuola
               Normale Superiore of Pisa, I-56100 Pisa, Italy} \\
\r {28} {\eightit University of Pittsburgh, Pittsburgh, Pennsylvania 15260} \\
\r {29} {\eightit Purdue University, West Lafayette, Indiana 47907} \\
\r {30} {\eightit University of Rochester, Rochester, New York 14627} \\
\r {31} {\eightit Rockefeller University, New York, New York 10021} \\
\r {32} {\eightit Rutgers University, Piscataway, New Jersey 08855} \\
\r {33} {\eightit Academia Sinica, Taipei, Taiwan 11530, Republic of China} \\
\r {34} {\eightit Texas A\&M University, College Station, Texas 77843} \\
\r {35} {\eightit Texas Tech University, Lubbock, Texas 79409} \\
\r {36} {\eightit Istituto Nazionale di Fisica Nucleare, University of Trieste/
Udine, Italy} \\
\r {37} {\eightit University of Tsukuba, Tsukuba, Ibaraki 315, Japan} \\
\r {38} {\eightit Tufts University, Medford, Massachusetts 02155} \\
\r {39} {\eightit Waseda University, Tokyo 169, Japan} \\
\r {40} {\eightit University of Wisconsin, Madison, Wisconsin 53706} \\
\r {41} {\eightit Yale University, New Haven, Connecticut 06520} \\
\end{center}
}

\date{\today}
 


\maketitle

\begin{abstract}
We present a measurement of
the time-dependent asymmetry in the rate for
$\overline{B}{^0_d}$ versus $B^0_d$ decays to $J/\psi K^0_S$.
In the context of the Standard Model this is interpreted as a
measurement of the $CP$-violation parameter $\sin(2\beta)$.
A total of $198\pm 17$ $B^0_d/\overline{B}{^0_d}$ decays were 
observed in $p\bar{p}$ collisions at $\sqrt{s} = 1.8$ TeV
by the CDF detector at the Fermilab Tevatron.
The initial $b$-flavor is determined by a same side flavor tagging technique.
Our analysis results in 
$\sin(2\beta) = 1.8\pm 1.1{\rm (stat)}\pm 0.3{\rm (syst)}$.
\end{abstract}

\narrowtext

The origin of Charge-Conjugation--Parity ($CP$) non-conservation in weak 
interactions has been an outstanding question in physics since its 
unexpected discovery in $K^0_L \rightarrow \pi^+\pi^-$ decays 
in 1964~\cite{CP64}. The favored mechanism for explaining $CP$ violation
is through the relationship between the weak interaction and the mass
eigenstates of quarks, which is described in the Standard Model (SM)
by the Cabibbo-Kobayashi-Maskawa (CKM) mixing matrix~\cite{CKM}.
With the addition of the third generation of quarks,
top and bottom, this matrix gains a physical complex
phase capable of explaining $CP$ violation.

After more than three decades, the $K^0$ remains the
only system where $CP$ violation has been observed.
Searches for $CP$ violation have recently been extended to
inclusive $B$ meson decays. However, the 
effects are expected to be small ($\sim\!10^{-3}$),
and no measurement has had the precision to 
reveal an effect~\cite{CPB}.

$CP$ violation is expected to have a large effect 
in the relative decay rate
of $B^0_d$ and $\overline{B}{^0_d}$ to the $CP$ eigenstate
$J/\psi K^0_S$~\cite{kshort}.
The interference of direct 
decays ($B^0_d \rightarrow J/\psi K^0_S$) {vs.}~those that have undergone 
mixing ($B^0_d \rightarrow \overline{B}{^0_d} \rightarrow J/\psi K^0_S$)
gives rise to a decay asymmetry 
\begin{equation}
{\cal A}_{CP}(t) \equiv \frac{\overline{B}{^0_d}(t)-B^0_d(t) }
                        {\overline{B}{^0_d}(t)+B^0_d(t) } = 
                       \sin(2\beta) \sin (\Delta m_d t),
\label{eq:cp_asym}
\end{equation}
where $B^0_d(t)$ ($\overline{B}{^0_d}(t)$) is the number of decays
to $J/\psi K^0_S$ at proper time $t$ given that the 
produced meson (at $t=0$) was a $B^0_d$ ($\overline{B}^0_d$).
The $CP$ phase difference between the two decay paths
appears via the factor $\sin(2\beta)$, and the
flavor oscillation through the mass difference $\Delta m_d$ between
the two  $B^0_d$ mass eigenstates.
Within the SM, constraints on the CKM matrix 
imply $0.30\leq\sin(2\beta)\leq 0.88$ at 95\% C.L.~\cite{Ali}.
Using a $J/\psi K^0_S$ sample, the OPAL Collaboration has recently reported
$\sin(2\beta) = 3.2 ^{+1.8}_{-2.0} \pm 0.5$~\cite{OPAL}.

Here we report on an analysis using
$B^0_d/\overline{B}^0_d\rightarrow J/\psi K^0_S$ decays
extracted from a 110 pb$^{-1}$ data sample of $p\bar{p}$ collisions
at $\sqrt{s}=1.8$ TeV collected in 1992-96 
by the CDF detector at the Fermilab Tevatron collider.
A description of the CDF detector may be found 
in Refs.~\cite{CDFDET,SVX}. 

The $B^0_d/\overline{B}{^0_d} \rightarrow J/\psi K^0_S$ 
sample selection~\cite{Ken} closely parallels Ref.~\cite{PRD}.
The $J/\psi$ is reconstructed via the $\mu^+\mu^-$ mode.
Both muons must be measured by our silicon vertex detector (SVX)~\cite{SVX}, 
thereby providing a precise decay length measurement. 
$K^0_S$ candidates are sought by fitting pairs of oppositely charged tracks,
assumed to be pions,
to the $K^0_S  \rightarrow \pi^+\pi^-$ hypothesis.
The $J/\psi$ and $K^0_S$ daughter tracks are combined in a four particle fit 
assuming they arise from
$B^0_d/\overline{B}{^0_d} \rightarrow J/\psi K^0_S$: 
the  $\mu^+\mu^-$ and $\pi^+\pi^-$
are constrained to their parents' world average
masses and separate decay vertices, 
and the $K^0_S$, $J/\psi$, and $B$ are
constrained to point back to their points of origin. 
A $B$ candidate is accepted if its transverse momentum with respect
to the beam line $p_T(B)$ is greater than 4.5 GeV/$c$, and 
if the $K^0_S$ candidate has $p_T(K^0_S) > 0.7$ GeV/$c$
and a decay vertex significantly displaced from the $J/\psi$ vertex.
Fit quality criteria are also applied.

\begin{figure}
\centerline{
\epsfysize 7.2cm
\epsfbox{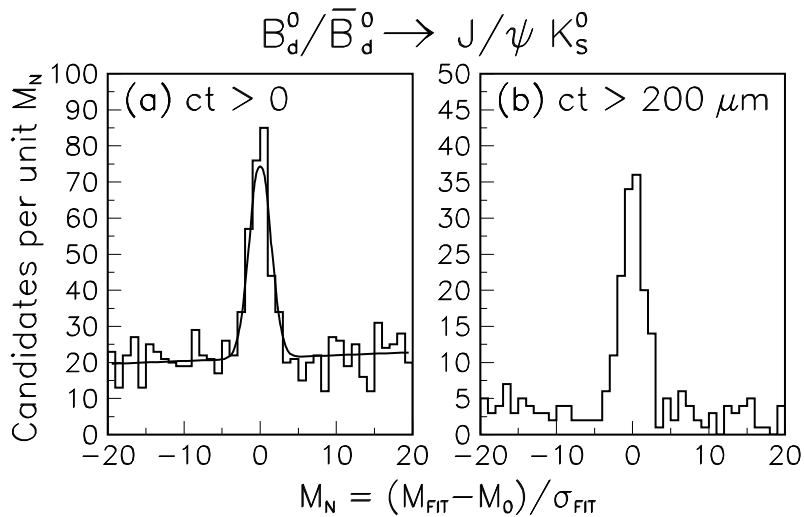} }
\caption{The normalized mass distribution of the $J/\psi K^0_S$ 
candidates with $ct > 0$ and $200 \, \mu$m. The curve is the Gaussian 
signal plus linear background from the likelihood fit (see text).
}
\label{fig:PsiKMass}
\end{figure}

We define $M_N \equiv (M_{FIT} - M_0)/\sigma_{FIT}$, where $M_{FIT}$ is the
mass of the $B$ candidate from the fit described above, 
$\sigma_{FIT}$ is its uncertainty (typically $\sim\!9\;{\rm MeV}/c^2$),
and $M_0$ is the central value of the $B^0_d$ mass peak.
The decay length of the $B$ 
is used to calculate its proper 
decay length $ct$, which includes the sign from the scalar product
of the transverse components of the vectors for the
$B$ decay vertex displacement from the $p\bar{p}$ interaction
vertex and the $B$ momentum.
The normalized masses  $M_N$ 
for the accepted candidates with $ct >0$
are shown in Fig.~\ref{fig:PsiKMass}a along with
the results of the likelihood fit described later.
The fit yields (for all $ct$) $198 \pm 17$ $B^0_d/\overline{B}{^0_d}$ mesons.

Measuring ${\cal A}_{CP}(t)$
is predicated upon
knowing whether the production ``flavor'' of the meson
was $B^0_d$ or $\overline{B}{^0_d}$. 
We determine this by a same side tagging (SST) method which
relies upon the correlation between the $B$ flavor and
the charge of a nearby particle. 
Such a correlation can arise from the fragmentation processes 
which form a $B$ meson from a $\bar{b}$ quark,
as well as from the pion from the decay of 
$B^{**}$ mesons~\cite{Rosner}. In both cases a $B^0_d$ 
is preferentially associated with a positive
particle, and a $\overline{B}{^0_d}$ with a negative one. 
The effectiveness of this method has been demonstrated by tagging 
$B \rightarrow \nu\ell D^{(*)}$ decays and observing
the time dependence of the $B^0_d$-$\overline{B}{^0_d}$ oscillation
and  measuring  $\Delta m_d$.
We have also measured the amplitude of the oscillation
({\it i.e.}, the strength of the correlation)
in a lower-statistics $B^0_d \rightarrow J/\psi K^{*0}$ sample and
found it to be consistent with 
the $\nu\ell D^{(*)}$ data~\cite{PRD,PRL}.

Our  SST method, following Ref.~\cite{PRL}, selects a 
single charged particle as a flavor tag 
from those within 
an $\eta$-$\phi$ cone of half-angle 0.7 around the $B$ direction, where
$\eta \equiv -\ln [\tan (\theta/2)]$ is the pseudorapidity,
$\theta$ is the polar angle relative to the outgoing proton beam
direction, and $\phi$ is the azimuthal angle around the beam line.
The tag must have 
$p_T >400$ MeV/$c$ and come from the $p\bar{p}$ interaction 
vertex ({\it i.e.}, 
have a transverse impact parameter within 
3 standard deviations of the interaction vertex).
If there is more than one candidate, the one  with the smallest
$p_T^{rel}$ is selected as {\it the} flavor tag, where 
the $p_T^{rel}$ of a particle is the component of its
momentum transverse to the momentum of the combined $B+$particle system.

We apply the SST method to the $J/\psi K^0_S$ sample. 
The tagging efficiency is $\sim\!65\%$.
The breakdown of tags is given in Table~\ref{tab:tags}
in proper time bins.
We call $|M_N|<3$ the ``signal region'' 
and $3<|M_N|<20$ the ``sidebands.''
Since negative (positive) tags are associated 
with $\overline{B}{^0_d}$'s ($B^0_d$'s), we form the 
asymmetry 
\begin{equation}
{\cal A}(ct) \equiv \frac{N^-(ct)-N^+(ct) }
                         {N^-(ct)+N^+(ct) }
\label{eq:meas_asym}
\end{equation}
analogous to Eq.~(\ref{eq:cp_asym}),
where $N^\pm(ct)$ are the numbers of positive and negative tags in 
a given $ct$ bin. 
The signal events generally have a positive
asymmetry ({\it i.e.}, favoring negative tags) at large $ct$.
The sidebands show a consistent negative asymmetry
(positive tags), but this has a small effect in the 
sideband-subtracted asymmetry at larger $ct$, where the signal
purity is high (see Fig.~\ref{fig:PsiKMass}b).

\begin{figure}
\centerline{
\epsfysize 8.6cm
\epsfbox{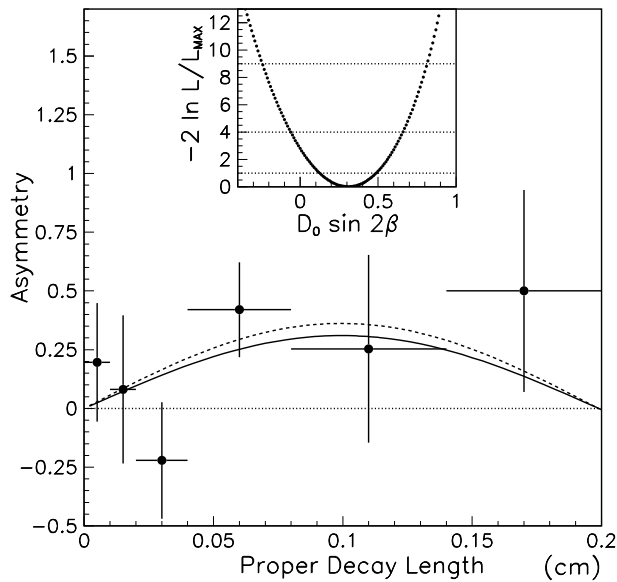}}
\caption{The sideband-subtracted tagging asymmetry as a function 
of the reconstructed  $J/\psi K^{0}_S$ proper decay length  (points).
The dashed curve is the result of a simple $\chi^2$ fit to 
${\cal A}_0\sin(\Delta m_d t)$.
The solid curve is the likelihood fit result, and the inset shows a
scan through the log-likelihood function as ${\cal D}_0\sin(2\beta)$ is
varied about the best fit value.
}
\label{fig:PsiKAsym}
\end{figure}

The sideband-subtracted asymmetries of Table~\ref{tab:tags} are displayed
in Fig.~\ref{fig:PsiKAsym} along with a $\chi^2$ fit (dashed curve) to
${\cal A}_0\sin(\Delta m_d t$), where $\Delta m_d$ is fixed to 
$0.474 \;{\rm ps}^{-1}$~\cite{PDG}.  
The amplitude, ${\cal A}_0 = 0.36\pm 0.19$, 
measures $\sin(2\beta)$ attenuated by a ``dilution factor'' 
${\cal D}_0\equiv 2{\cal P}_0-1$, where ${\cal P}_0$ is the probability
that the tag correctly identifies the $B^0_d$ flavor.
The determination of ${\cal A}_0$ is dominated by the asymmetries
at larger $ct$'s due to the $\sin(\Delta m_d t)$ shape;
this is also where the background is very low.

We refine the fit
using an unbinned maximum likelihood fit based
on Ref.~\cite{PRD}.  This fit makes optimal use of the low 
statistics by fitting signal and background distributions in $M_N$ and $ct$,
including sideband and $ct<0$ events which help constrain the background.
The likelihood fit also incorporates resolution effects and
corrections for systematic biases, such as the
inherent charge asymmetry favoring positive tracks
resulting from the
wire plane orientation in the main drift chamber.  

We measure the intrinsic charge asymmetry  of the tagging 
in a large inclusive (unflavored) $J/\psi$ sample with displaced decay
vertices ($>90\%$ $b$ hadrons) and parameterize its dependence on
track $p_T$ and event occupancy.  The occupancy dependence is weak.  At
$400\;{\rm MeV}/c$, the SST $p_T$ threshold, the asymmetry is
$5.6\pm 1.1\%$, falling as $p_T^{-4}$ to
$0.14\pm 0.86\%$ at high $p_T$ 
(the average tag asymmetry in the $J/\psi$ sample is $1.6\pm0.7$\%), 
all favoring positive tags.  This correction is applied to the
signal in the likelihood fit; the charge asymmetry 
of the $J/\psi K^0_S$ background
is measured independently by the fit itself.

The solid curve in Fig.~\ref{fig:PsiKAsym} is the result of the likelihood
fit, which gives 
${\cal D}_0\sin(2\beta)=0.31\pm 0.18$.  
As expected, the
two fits give similar results, indicating that our result is dominated
by the sample size and that the corrections and improvements of the
likelihood fit introduce no dramatic effects.
Also shown in the Fig.~\ref{fig:PsiKAsym} inset
is the relative log-likelihood as a function of ${\cal D}_0\sin(2\beta)$;
the shape is parabolic, indicating Gaussian errors.

As noted above, the sidebands favor positive tags.
The maximized likelihood ascribes an asymmetry of $16.7 \pm 8.2$\%,
or an $\sim\!2\sigma$ excess of positive tags, 
to the long-lived backgrounds 
({\it e.g.}, $B\rightarrow J/\psi X$ with an unassociated $K_S^0$).
Prompt background (consistent with 
$ct$-resolution) has an asymmetry of $0.6 \pm 4.5$\%
favoring negative tags.

Systematic effects from $B$ backgrounds have been considered.
For instance, the decay $B^0_d \rightarrow J/\psi K^{*0}$,
$K^{*0} \rightarrow K^0_S \pi^0$, where we do not reconstruct
the $\pi^0$, has a negligible effect on the result.
Background asymmetries were also studied in $J/\psi K^+$ and $J/\psi K^{*0}$
modes~\cite{PRD}.  No systematic pattern emerged.  Since no other
biases have been found aside from the above small effects,
we attribute the background asymmetry largely to statistical fluctuations.
Again, the effect of these asymmetries is small as the total background
fraction is small at large $ct$ (see Fig.~\ref{fig:PsiKMass}b)
where $\sin (\Delta m_d t)$ is large.

We determine the systematic uncertainty on ${\cal D}_0\sin(2\beta)$
by shifting the central value of each fixed input
parameter to the fit by $\pm 1\sigma$ and refitting to find the
shift in ${\cal D}_0\sin(2\beta)$.
Varying the $B^0_d$ lifetime ($468\pm 18\;\mu{\rm m}$~\cite{PDG})
shifts the central value by $\pm 0.001$.
The parameterization of the intrinsic charge asymmetry 
is also varied, yielding a
$^{+0.016}_{-0.019}$ uncertainty.  
The largest systematic uncertainty is
due to $\Delta m_d = 0.474\pm 0.031\;{\rm ps}^{-1}$~\cite{PDG}, 
which gives a $^{+0.029}_{-0.025}$ shift.  
These systematic uncertainties are added in quadrature, 
giving  ${\cal D}_0\sin(2\beta) = 0.31 \pm 0.18\pm 0.03$.

To obtain $\sin(2\beta)$, we use dilution measurements 
from other $B$ samples.
Our best single ${\cal D}_0$  measurement,
from a large $B \rightarrow \ell D^{(*)}X$ sample,
is $0.181 ^{+0.036}_{-0.032}$~\cite{PRD,PRL}. 
Because of differing lepton
$p_T$ trigger thresholds, the average $p_T$ of the 
semileptonic $B$'s is $\sim\!21$ GeV/$c$, but it is only  $12$ GeV/$c$
in the $J/\psi K^0_S$ data. 
We correct for this difference by using a version 
of the PYTHIA event generator~\cite{PYTHIA} 
tuned to CDF data~\cite{PRD,Dejan}.
We supplement the above measurement 
of ${\cal D}_0$ with the dilution  ${\cal D}_+$
measured from $B^+$'s in
the same $\ell D^{(*)}$ sample, as well as 
measurements 
from $B\rightarrow J/\psi K^+$ and $J/\psi K^{*0}$.  
The simulation also accounts for the systematic difference between 
$B^0_d$ and $B^+$ dilutions.
The ${\cal D}_0$ appropriate for our
$J/\psi K_{S}^{0}$ sample is then $0.166\pm 0.018\pm 0.013$, a small
shift from 
0.181.
The first error is due to the uncertainty in the dilution measurements,
and the second  is due to the Monte Carlo  extrapolation. 
The latter is determined by 
surveying a range of simulation parameters~\cite{Ken,PRD}.

Using this ${\cal D}_0$, we find that
$\sin(2\beta)=1.8\pm 1.1\pm 0.3$.
The central value is  unphysical
since the amplitude of the measured 
asymmetry is larger than  ${\cal D}_0$.
If one wishes to frame this result in terms of confidence
intervals, various alternatives are available~\cite{PDG,Cousins}.
We follow the frequentist construction of Ref.~\cite{Cousins},
which gives proper confidence intervals even for
measurements in the unphysical region.
Our measurement thereby corresponds to excluding $\sin(2\beta)<-0.20$
at a 95\% confidence level (C.L.).
We also calculate that if the true value of $\sin(2\beta)$ were 1,
the median expectation of an exclusion for an experiment like ours would
be $\sin(2\beta)<-0.89$ at 95\% C.L.  This is a
measure of experimental sensitivity~\cite{Cousins};
our limit is higher, reflecting the excursion into
the unphysical region.

It is interesting to note that if ${\cal D}_0 \not = 0$,
the exclusion of $\sin(2\beta)=0$ is {\it independent}
of the value of ${\cal D}_0$. Given ${\cal D}_0 > 0$,
the same prescription as above yields a dilution-independent
exclusion of $\sin(2\beta) < 0$ at 90\%  C.L.

We have explored the robustness of our result by varying
selection and tagging criteria. 
None had a significant effect on the asymmetry with the exception 
of the tagging  $p_T$ threshold.
In principle, any choice of the
threshold would give an unbiased estimator of ${\cal D}_0\sin(2\beta)$.
The 400 MeV/$c$ threshold, however, was our {\em a priori} choice,
taking into account the tracking asymmetry at
lower $p_T$ and the reduced tagging efficiency at higher $p_T$.

When varying the $p_T$ threshold 
we found that
${\cal D}_0\sin(2\beta)$  drops rather sharply in going  
from 0.5 to 0.6 GeV/$c$, and then gradually rises.
The probability of observing such a large change in an $\sim\!100$ MeV/$c$
step is estimated to be $\sim\!5\%$.
The smallest value of ${\cal D}_0\sin(2\beta)$ is 
$-0.12 \pm 0.21$ (stat. error only) 
for a 0.7 GeV/$c$ threshold.  This variation cannot be attributed 
to the dependence of ${\cal D}_0$ on the $p_T$
threshold: both the charged and neutral dilution measurements
vary slowly, in good agreement with the PYTHIA calculations~\cite{PRD}.
Moreover, no systematic effects have been found which 
are able to account for
such a variation.

As we can identify no mechanism to give the particular behavior seen,
we characterize the 
variation of ${\cal D}_0\sin(2\beta)$ with the $p_{T}$ threshold
by calculating the probability
that the
variation in the data
agrees with the 
slow variation in the simulation.
To this end we employ the $\chi^2$ procedure used in
Ref.~\cite{PRD} to study the dilution variation in a 
$B^+\to J/\psi K^+$ sample.
We compare the data with Monte Carlo pseudo-experiments of
similar size and find that the probability of obtaining a higher
$\chi^2$ ({\em worse} agreement) than the data is 42\%, considering
only statistical fluctuations~\cite{Ken}.
Thus, the observed variation of ${\cal D}_0\sin(2\beta)$
with the SST $p_T$ threshold is consistent with statistical fluctuations 
expected for a sample of this size.

In summary, we have applied a same side flavor tagging method 
to a sample of $B^0_d/\overline{B}{^0_d} \rightarrow J/\psi K^0_S$ decays
and measured 
$\sin(2\beta) = 1.8\pm 1.1\pm 0.3$.
Although the sensitivity of the result on the tagging $p_T$
threshold complicates the interpretation, 
our result favors current Standard Model expectations 
of a positive value of $\sin(2\beta)$.

This result establishes
the feasibility of measuring $CP$ asymmetries in $B$ meson decays at a hadron
collider. Operation of the Main Injector in the next
Tevatron Collider run should provide more than an order of magnitude increase
in luminosity.
Detector upgrades will further enlarge our $B$ 
samples. If current expectations are correct,
these large samples should be
sufficient to observe and study $CP$ violation in $J/\psi K^0_S$,
and possibly in other modes as well~\cite{CDFII}.

     We thank the Fermilab staff and the technical staffs of the
participating institutions for their vital contributions.  This work was
supported by the U.S. Department of Energy and National Science Foundation;
the Italian Istituto Nazionale di Fisica Nucleare; the Ministry of Education,
Science and Culture of Japan; the Natural Sciences and Engineering Research
Council of Canada; the National Science Council of the Republic of China; 
the Swiss National Science Foundation; and the A. P. Sloan Foundation.

\narrowtext
\begin{table}
\begin{center}
\begin{tabular}{ccccccccc}   
$ct$ & \multicolumn{3}{c}{Signal} & $\,$ &\multicolumn{3}{c}{Sidebands} 
     & Asymmetry \\ 
\cline{2-4} \cline{6-8}
($\mu$m)   &   $-$  &  $+$ &  $0$ &&  $-$ &  $+$ &  $0$    &  ($\%$) \\ \hline
$-200$ - 0 &    42  &   21 &  43  &&  167 &  193 &  174    & ---   \\
   0 - 100 &    53  &   48 &  49  &&  156 &  175 &  205    &$ 20 \pm 25$ \\
 100 - 200 &    14  &   14 &  15  &&   26 &   34 &   24    &$  8 \pm 32$ \\
 200 - 400 &    12  &   18 &  19  &&   17 &   22 &   10    &$-22 \pm 24$ \\
 400 - 800 &    26  &   13 &  22  &&   11 &   18 &   11    &$ 42 \pm 18$ \\
 800 - 1400&     6  &    4 &   9  &&    6 &    6 &    2    &$ 25 \pm 40$ \\
1400 - 2000&     3  &    1 &   1  &&    0 &    0 &    2    &$ 50 \pm 43$ \\
\end{tabular}
\end{center}
\caption{\small Tags for the $J/\psi K^0_S$ candidates in proper
decay length ($ct$) bins. The signal region is $|M_N| < 3 $,
and the sidebands are $3 < |M_N| < 20$. The ``$+$,'' ``$-$,'' and ``0''
headings are for positive, negative, and untagged events. The last column
is the sideband-subtracted tagging asymmetry 
[Eq.~(\ref{eq:meas_asym})].
The asymmetry for the background-dominated first row is not quoted
because there is not a tagged, sideband-subtracted excess.
}
\label{tab:tags}
\end{table}


\end{document}